# Time-Stretched Femtosecond Lidar using Microwave Photonic Signal Processing

Lijie Zhao, Haiyun Xia, Yihua Hu, Tengfei Wu, Zhen Zhang, Jibo Han, Yunbin Wu, and Tiancheng Luo

*Abstract*—A real-time lidar with 0.1 Mega Hertz update rate and few-micrometer resolution incorporating dispersive Fourier transformation and instantaneous microwave frequency measurement is proposed and demonstrated. As a femtosecond laser pulse passing through an all-fiber Mach-Zehnder Interferometer, where the detection light beam is inserted into the optical path of one arm, the displacement is encoded to the frequency variation of the temporal interferogram. To deal with the challenges in storage and real-time processing of the microwave pulse generated on a photodetector, we turn to microwave photonic signal processing. A carrier wave is modulated by the time-domain interferogram using an intensity modulator. After that, the frequency variation of the microwave pulse is uploaded to the first order sidebands. Finally, the frequency shift of the sidebands is turned into transmission change through a symmetric-locked frequency discriminator. In experiment, a real-time ranging system with adjustable dynamic range and detection sensitivity is realized by incorporating a programmable optical filter. Standard deviation of 7.64 μm, overall mean error of 19.10 μm over 15 mm detection range and standard deviation of 37.73 μm, overall mean error of 36.63 μm over 45 mm detection range are realized, respectively.

Manuscript received; revised; accepted. Date of publication; date of current version. This work was supported in part by the National Key Research and Development (R&D) Plan of China (2018YFB0504300) and supported by shanghai Municipal Science and Technology Major Project under Grant No. 2019SHZDZX01. (*Corresponding author: Haiyun Xia and Tengfei Wu*).

Lijie Zhao, Haiyun Xia, Zhen Zhang, Yunbin Wu and Tiancheng Luo are with Chinese Academy of Sciences Key Laboratory of Geospace Environment, School of Earth and Space Sciences, University of Science and Technology of China, Hefei, Anhui, 230026, China. (e-mail: zhaolj@mail.ustc.edu.cn; hsia@ustc.edu.cn; znzhang@mail.ustc.edu.cn; wuyunbin@mail.ustc.edu.cn; ltc8@mail.ustc.edu.cn).

Yihua Hu is with State Key Laboratory of Pulsed Power Laser Technology, National University of Defense Technology, Hefei 230037, China (e-mail: yh_hu@263.net).

Tengfei Wu and Jibo Han are with Science and Technology on Metrology and Calibration Laboratory, Changcheng Institute of Metrology and Measurement, Aviation Industry Corporation of China, Beijing 100095, China. (e-mail: tengfei.wu@163.com; linghanxiyue@163.com).

## I. INTRODUCTION

Femtosecond laser has been a promising tool for ultrafast and precise detection applications, such as high speed gas spectrum analysis [1]-[3], microscopic imaging [4] and ultrafast distance measurement [5]-[9], due to its high repetition rate, short exposure time, wide spectral range and high stability [10]. Ultrafast ranging with high resolution and high acquisition rate is attractive to ultrafast dynamic researches and high-speed industry tests. By introducing the dispersive Fourier transformation technology [11]-[13], single-shot ultrafast range detection can be realized at large dynamic range with nanosecond acquisition time and nanometer accuracy [14], [15]. The detection of fast flying target at speed of tens of kilometers per second has been reported [16]-[18]. However，the measured signal data size can be tremendous because a high sampling rate analog to digital converter (ADC) is usually needed. For oscilloscope with 80 GHz sampling rate and 8-bit vertical resolution, the data size can approach 150 GBs per second with binary data format. It is a great burden to achieve data storage and real-time processing continuously. Short time data storage and time-consuming post-processing hinders its implementations where continuous detection and real-time feedback is required.

In recent years, microwave photonics has shown immense potential to overcome the problems mentioned above and realize real-time signal processing with integrated functionality. Using microwave photonics methods, electrical domain signal from photodetector (PD) can be converted into optical domain signal by an electro-optic modulator and processed in optical links directly [19]. One of the focused topics in microwave photonics is instantaneous microwave frequency measurement and has been densely investigated in recent years. It is of great importance for high-resolution real-time applications. Fortunately, in the ultrafast lidar based on dispersive Fourier transformation, displacement is encoded into the frequency of time-domain interferogram which is a microwave pulse [20]-[23]. The frequency information can be processed instantaneously by microwave photonic approaches.

In direct-detection of optical frequency, there is a so-called edge technique in lidar applications. For example, in a Doppler lidar, Fabry–Perot interferometer (FPI) is used as an optical frequency discriminator. The Doppler shift of either the Rayleigh or the Mie backscattering signal is converted into its transmission change through the FPI [24]-[26]. Recently, using InGaAs/InP single photon detector (SPD) [27], [28], up-conversion SPD [29]-[31] or superconducting nanowire

SPD [32], [33], direct detection of Doppler shift at single photon level is achieved at 1.5 micrometer working wavelength.

For microwave frequency measurement, the same thought can be adopted. One commonly used scheme is by introducing a monotonous frequency-intensity mapping, namely amplitude comparison function (ACF). The ACF can be realized by different schemes, such as Sagnac loop [34], fiber Bragg grating [35], Echelle diffractive grating [36], birefringence effect in polarization-maintaining fiber (PMF) [37], a two-tap finite impulse response filter [38], photonic Brillouin filter [39], [40], frequency dependent DC output produced by optical mixing method [41], fade function generated by dispersive medium [42]-[46] or integrated photonic chips [47]-[50]. For multiple frequency measurement, methods such as frequency to time mapping [51], [52] and frequency shifter [53] are demonstrated. Other approaches can be found elsewhere [54], [55].

Here, we demonstrate a real-time range detection method based on microwave photonic signal processing, where the displacement is finally mapped to the transmission (signal intensity ratio) variance. The procedure is shown in Fig. 1.

Step1: the displacement is encoded into the frequency shift of microwave pulse by a technique called dispersive Fourier transform interferometer.

Step 2: the frequency shift of microwave pulse is uploaded into the 1$^{st}$ order sidebands of carrier wave under intensity electro-optic modulation.

Step 3: by adopting optical frequency direct-detection, the frequency shift of the above sidebands is converted into its transmission variance on a frequency discriminator.

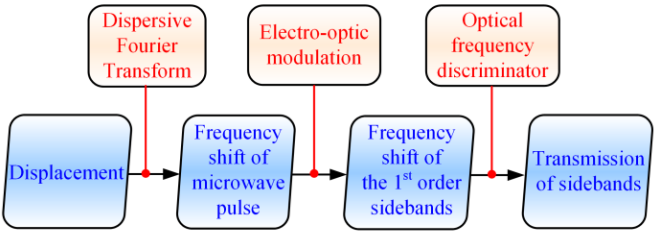

Fig. 1. Flowchart of real-time displacement measurement

## II. PRINCIPLE

The schematic diagram of the ultrafast femtosecond ranging system with microwave photonic signal processing is shown in Fig. 2. The displacement measurement is based on dispersive Fourier transformation method. The optical source is a femtosecond laser (FSL), with its repetition rate locked via a closed-loop control. The femtosecond laser pulse is filtered, then coupled into a Mach-Zehnder interferometer (MZI), splitting into its detection arm and reference arm. The light in detection arm is transmitted through a circulator and a collimator to a retroreflector, the retroreflector is mounted on a nanometer linear positioning stage (NLPS) which is driven by a servo controlled Piezoelectric transducer (PZT). In the reference arm, an optical fiber delayer (OFD) is used to adjust time delay difference between the two arms. Then the light from the two arms are combined and followed by two stages of dispersion compensation fiber (DCF) and Erbium-doped fiber amplifier (EDFA). After time-stretching and chirped-pulse amplification, the time-domain interferogram is generated. The polarization controllers (PC1, PC2) and inline polarizer (ILP) are used to obtain high contrast ratio of the interferogram.

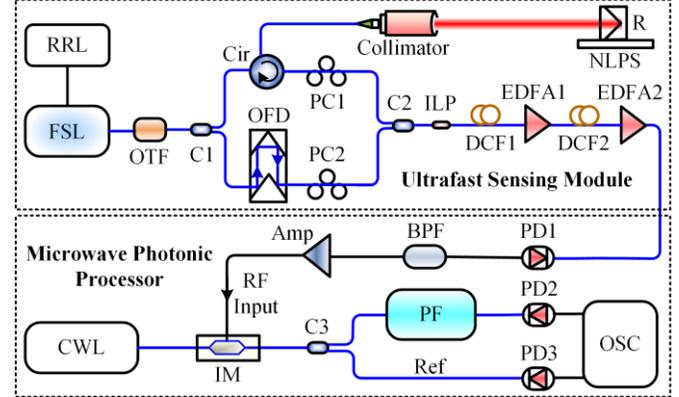

Fig. 2. Schematic diagram of ultrafast femtosecond laser ranging based on microwave photonic signal processing. RRL: Repetition Rate Locking system; FSL: Femtosecond laser; OTF: Optical tunable filter; C: Coupler; Cir: Circulator; PC: Polarization controller; ILP: In-line polarizer; OFD: Optical fiber delayer; R: Retroreflector; NLPS: Nanometer linear position stage; DCF: Dispersion compensation fiber; EDFA: Erbium-doped fiber amplifier; PD: Photodetector; BPF: Band-pass filter; Amp: Microwave amplifier; CWL: Carrier wave laser; IM: Intensity modulator; PF: Programmable filter; OSC: Oscilloscope

The interferogram is detected on a photodetector, yielding a microwave pulse train. A broadband electrical band-pass filter (BPF) is used to eliminate the direct current component and low frequencies disturbance such as repetition rate of laser and its harmonic components. A carrier wave laser is modulated by the amplified microwave signal through an intensity modulator, operating at carrier suppression mode. Then the modulated light with two sidebands are generated and divided into two channels. One channel includes a programmable filter and the other one is set as power reference channel. The programmable filter is used as frequency discriminator. By changing the filter bandwidth, adjustable detection sensitivity (thus different dynamic range) can be realized. Note that, high sensitivity is achieved at sacrifice of detection range. Finally, the optical signals at two channels are detected at two low-bandwidth photodetectors and sent to an oscilloscope for signal processing.

In the first step, the displacement relative to reference point is mapped to the frequency of the microwave pulse. The principle of displacement detection proposed in our previous work is briefly recalled here [14], [15]. The MZI in the system adopts balanced structure by adjusting the optical fiber delayer. Displacement to be measured can be obtained by time of flight $\tau$, which is contained in time delay between the two arms of MZI. For fast detection, DCF is used to stretch the femtosecond pulse and generate time-domain interferogram which contains time delay information. Dispersive Fourier transformation of the fiber in the system builds a unique relationship between time delay and frequency change. With third-order dispersion considered, the instantaneous optical frequency of time-stretched pulse can be written as [15], [23]:

$$f_{ins}(t) = f_0 + \frac{t}{2\pi\beta_2 L} - \frac{\beta_3 L t^2}{4\pi(\beta_2 L)^3} \quad (1)$$

where $f_0$ is center frequency, $\beta_2$ and $\beta_3$ are second-order and third-order dispersion coefficients of DCF, L denotes the fiber length. Then the instantaneous frequency of time-domain interferogram is the beat frequency of the pulses transmit in the two arms of the MZI with a time delay $\tau$. The time-domain interferogram detected by photodetector can be expressed by:

$$I_{TDI}(t) \propto a_m(t)\cos\{[f_{ins}(t+\tau) - f_{ins}(t)]t\}$$
$$= a_m(t)\cos\left\{\left[\frac{\tau}{\beta_2 L} - \frac{\beta_3 L \tau^2}{2(\beta_2 L)^3} - \frac{\beta_3 L \tau t}{(\beta_2 L)^3}\right]t\right\} \quad (2)$$

where $a_m(t)$ is normalized signal envelope. It is shown that the time delay which corresponding to the displacement is mapped to the microwave frequency of time-domain interferogram.

The displacement measurement dynamic range of proposed system is $c\tau_{Max}/2$, $\tau_{Max}$ is the maximum time delay and can be estimated by [14]:

$$\tau_{Max} = \frac{DL\lambda^2 \Delta f}{c} \quad (3)$$

where $D = -2\pi c\beta_2/\lambda^2$ is dispersion coefficient and $\lambda$ is center wavelength of optical pulse, $\Delta f$ is Optical-Electro (OE) bandwidth which is limited by the minimum bandwidth of PD1, bandpass filter or modulator. Nowadays, commercial devices with bandwidth exceed 100 GHz are available. According to (3), dynamic range can be enhanced by increasing dispersion or OE bandwidth. With large dispersion, optical power compensation is usually needed [20], [21]. It is worth noting that the time-stretched pulse should not overlap in time domain, thus the maximum round-trip time of detection pulse relative to reference pulse should be slightly smaller than $1/f_{rep}$ to avoid interference between neighboring pulses, where $f_{rep}$ is repetition rate of pulses.

In previous works, signal processing of time-domain interferogram is complex. Several mathematical manipulations are needed, including time-frequency transformation, interpolation, resampling in the frequency domain, inverse Fourier transformation and curve fitting. Here in order to simplify the processing, microwave photonics approach is adopted.

Hence in the second step, the microwave frequency change is treated by microwave photonic signal processor. The principle of signal processing is shown in Fig. 3. The microwave pulse train can be expressed by a convolution of a delta function $\sum \delta(t-nT)$ and microwave signal $u(t) = V_m a_m(t)\cos(2\pi f_m t)$, where $T$ is pulse period, $V_m$ and $f_m$ are microwave amplitude and frequency, respectively. The optical field of modulated light in single pulse period is given by [56]:

$$E(t) = \sqrt{P_0}\exp(i2\pi f_c t)\sin[\beta a_m(t)\cos(2\pi f_m t)] \quad (4)$$

where $P_0$ is input optical power, $f_c$ is optical carrier frequency, $\beta = \pi V_m/V_\pi$ is modulation depth, $V_\pi$ is half-wave voltage of intensity modulator.

The intensity modulator is also a Mach-Zehnder modulator (MZM) and biased at its minimum transmission point to suppress the power of the optical carrier. In practice, most MZMs are non-ideal and suffer from modulation chirp [57]. However, with small signal modulation regime, the chirp parameter of the modulator is fixed with time and independent of the power of microwave signal [58]. The modulation chirp effect can be canceled by two-channel detection with one channel set as the power reference. Furthermore, small signal modulation can suppress higher order modulation sidebands effectively. The optical field at small modulation condition can be written as:

$$E(t) \approx \frac{\sqrt{P_0}\beta}{2}a_m(t)\{\exp[i2\pi(f_c+f_m)t] + \exp[i2\pi(f_c-f_m)t]\} \quad (5)$$

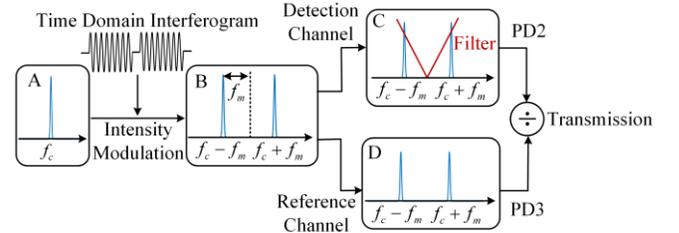

Fig. 3. Operation principle of microwave photonic signal processing

It shows that two sidebands are generated with equal frequency separation relative to carrier frequency. That means the frequency information of the microwave signal is uploaded to the sidebands. The two sidebands are then divided into two channels for transmission detection. A programmable filter is locked symmetrically, by setting its minimum transmission point at the frequency of the carrier wave. Then the transmission of the sidebands varies versus the frequency shift of microwave pulse.

The transmission response of detection channel can be written as:

$$T(f) = D(f) \otimes A_m^2(f) \quad (6)$$

Where $D(f)$ and $A_m(f)$ are designed filter function and Fourier transformation of $a_m(t)$ respectively.

For low bandwidth PDs, only the envelop of modulated signal can be captured, thus the output of detection and reference channel can be written as:

$$I_{det}(t) = \frac{c_1 \Re_1 P_0 \beta^2}{2}a_m^2(t)T(f_c+f_m) \quad (7)$$

$$I_{ref}(t) = \frac{c_2 \Re_2 P_0 \beta^2}{2}a_m^2(t) \quad (8)$$

where $c_1$, $c_2$ are coupling coefficients of two channels and $\Re_1$, $\Re_2$ are responsivity of two PDs. It is shown that the frequency of microwave signal fed to the modulator is mapped to optical intensity through optical filter in detection channel. Thus, the transmission function, so-called ACF of detection channel and reference channel can be obtained by $r(f_m) = kT(f_c+f_m)$, where $k = c_1\Re_1/c_2\Re_2$. Taking the ratio operation can eliminate the influence of optical power jitter and modulation depth fluctuation during the measurement.

In the third step, the displacement is mapped to a transmission function $r(f_m)$ which is designed to be monotonic. When the displacement varies, time delay of MZI in the sensing module and the frequency of time-domain interferogram varies accordingly. Thus, the transmission in the microwave photonic processor will vary along with displacement. Finally, the mapping of displacement to a transmission function is established by calibration.

The higher order dispersion effect induced by DCF used in the system and frequency response of the devices used in the microwave photonic signal processing module can be compensated in the calibration process. With a calibrated line of $r(f_m)$, the measured transmission data can be used to retrieve the displacement information in real-time.

## III. EXPERIMENT

In experiment, a homemade femtosecond laser (CIMM, TC1550-2G) has average output power of 30 mW and pulse width of about 93 fs. The pulse repetition rate is locked to 50 MHz. A small portion of the femtosecond laser pulse is filtered out with spectrum width of 8.2 nm centered at 1553 nm. The NLPS (PI, N-565.260) has a positioning resolution of 0.5 nm and bidirectional repeatability of 50 nm with a closed-loop control (PI-E861.1A1). The total dispersion used with two spools of DCF (ofs, DCM(D)-C-G.652-DCF) is -2298 ps/nm. To guarantee the signal to noise ratio, the total optical amplification factor of time-stretched pulse is 45.8 dB by using two cascaded EDFAs (Amonics, AEDFA-PA-35). The PD1 (Alphalas, UPD-15-IR2-FC) used for pulsed microwave detection is an InGaAs ultrafast photodetector with 25 GHz bandwidth. The electric BPF (Wainwright, WHNX6) used before modulation has passband of 2.3-26.5 GHz. A continuous wave laser (Keyopsys, PEFL-EOLA) with center wavelength of 1548.5 nm is used as optical carrier. By tuning the time delayer, reference point of the Retroreflector (zero position) is settled, when the center frequency of the microwave pulse is 2.3 GHz. Since the bandwidth of intensity modulator (Photline, MXER-LN) is 20 GHz, thus the OE bandwidth is 20 GHz. The microwave pulse can be processed continuously, with a span from 2.3 GHz to 20 GHz. The maximum displacement measurement dynamic range is 4.87 cm according to (3). The bandwidth of PD2 and PD3 (Thorlabs, PDB430C) is 350 MHz. Key parameters of the system are listed in table Ⅰ for reader's convenience.

To avoid effects of environment fluctuations and vibrations during measurement, room temperature is set about 24 ℃ and all fiber devices and positioning stage are mounted on an air floating platform. The intensity modulator is placed in a thermotank at 24 ℃ with TEC control. Its temperature fluctuation is less than 0.01 ℃.

Several specific displacements to be measured are chosen within the dynamic range. The microwave pulse is captured by an Oscilloscope (Teledyne Lecroy, LabMaster MCM-Zi-A) with 80 GHz sampling rate. As shown in Fig. 4, three waveforms at different displacements (5 mm, 15 mm, 25 mm) are plotted. One can see the obvious frequency change related to the displacement.

TABLE I
SUMMARY OF SYSTEM PARAMETERS

| Parameters | Values |
|---|---|
| **Sensing Module:** | |
| Femtosecond laser: | |
| Center wavelength | 1560 nm |
| Repetition rate | 50 MHz |
| Pulse width | 93 fs |
| Average power | 30 mW |
| Collimated beam diameter | 3.6 mm |
| Total dispersion of DCF | -2298 ps/nm |
| Total optical amplification | 45.8 dB |
| Bidirectional repeatability of NLPS | 50 nm |
| **Microwave Photonic Processor:** | |
| Passband of electric BPF | 2.3-26.5 GHz |
| Microwave amplification | 30 dB |
| Carrier wave laser wavelength | 1548.495 nm |
| Bandwidth of intensity modulator | 20 GHz |
| Bandwidth of PD1 | 25 GHz |
| Bandwidth of PD2, PD3 | 350 MHz |

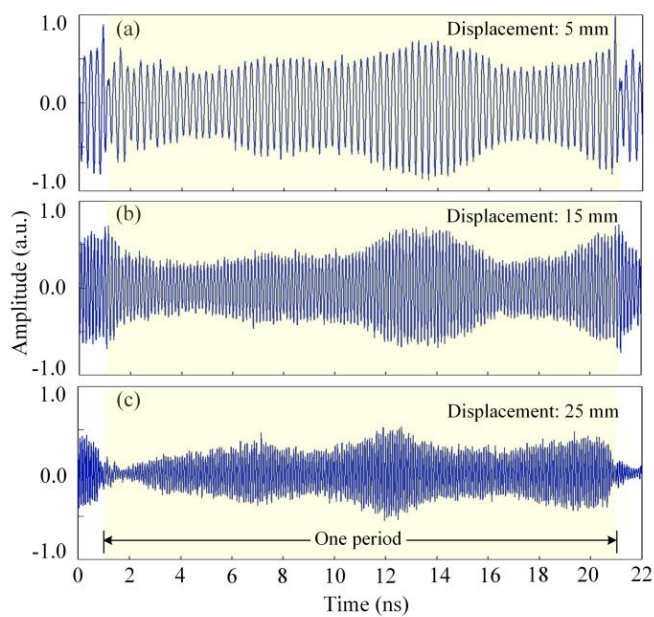

Fig. 4. Microwave pulse waveform with specific displacements

The microwave pulse train is then amplified 30 dB by a broadband microwave amplifier (SHF, L806A) and sent to the intensity modulator. As shown in Fig. 5, the sidebands corresponding to different center frequencies of the microwave pulses are plotted, which are measured by an optical spectrum analyzer (OSA, Yokogawa, AQ6370C). The frequency is defined relative to the absolute optical frequency (193.6025 THz) of the carrier wave. The position of the two sidebands are symmetrically located about the carrier frequency and shift away from the center along with increased displacement.

With double sidebands carrier suppression modulation, the power of sidebands can achieve 30 dB higher than the power of carrier. As displacement increases, the coupling efficiency of the collimator from space to the polarization-maintaining fiber decays, resulting a weaker microwave power fed to the modulator. So the intensity of the sidebands drops down.

The modulated signal is divided into two channels for transmission detection. In the detection channel, filter functions are specifically designed to be a symmetrical linear function relative to carrier frequency. Thus the dynamic range of displacement detection can be adjusted by setting the bandwidth. There is a trade-off between dynamic range and detection sensitivity. Higher detection sensitivity can be obtained at the expense of dynamic range by decreasing the bandwidth of the filter.

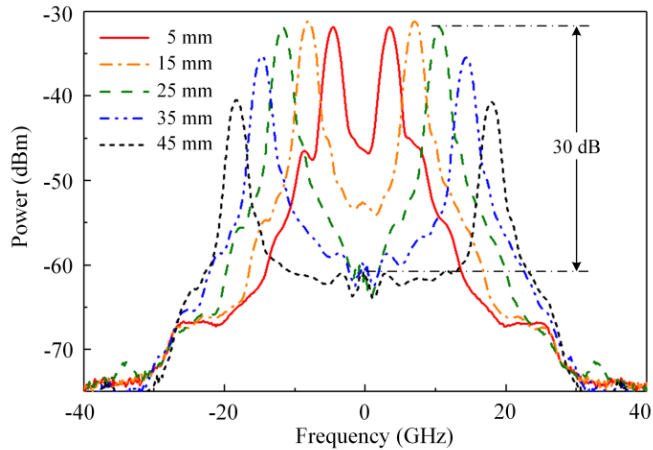

Fig. 5. Spectrum of modulated signal with specific displacements

In experiment, two filters with different detection sensitivity are designed. Dynamic range of filter 1 is designed to be 15 mm and filter 2 is 45 mm. The programmable filter used here is WaveShaper (FINISAR, 1000A), the input optical signal is first dispersed by a conventional diffraction grating and then processed by a Liquid Crystal on Silicon (LCoS) optical processor. Thus, spectral attenuation profile can be designed. As shown in the Fig. 6, the frequency response of the two filters are measured by an OSA with an amplified spontaneous emission (ASE) source. The sweep resolution of the OSA is set to 2.5 GHz. Thus, the measured filter response curves broaden due to the convolution of response of the filters and the OSA. Through the filters, a monotonic transmission function versus frequency shift of sidebands is established. The intensity signal from PDs in detection and reference channels are captured by an oscilloscope (Teledyne Lecroy, HDO6034).

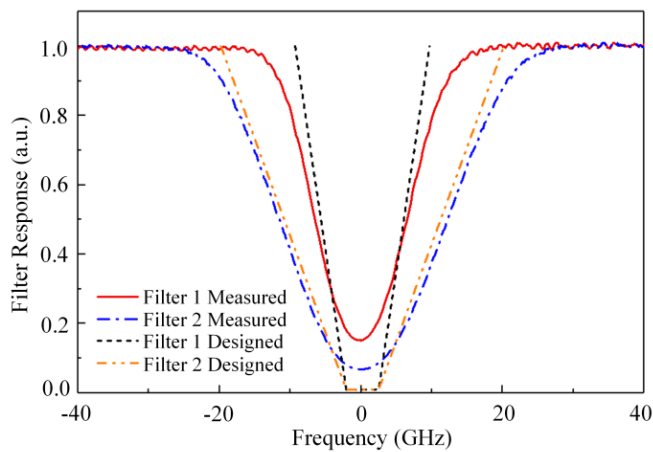

Fig. 6. Measured filter response with different sensitivity

Before the range measurement, the system requires careful calibration. At different positions of retroreflector controlled by NLPS, transmission values are measured. The detection signal and reference signal should be synchronized firstly in the time domain, by tuning the time delay between the two channels on the oscilloscope. Thus, the transmission can be calculated directly.

During the transmission measurement, the sampling rate of the oscilloscope is set to 1.25 GHz with 12-bit vertical resolution. The calibrated transmission line of two filters with different dynamic ranges by 3rd order polynomial fitting are shown in Fig. 7. The calibration equation can be expressed by:

$$y = ax + bx^2 + cx^3 + d \quad (9)$$

The coefficients of calibration equation for 15 mm and 45 mm are shown in table Ⅱ. With calibration line determined, corresponding displacement can be obtained immediately through measured transmission data.

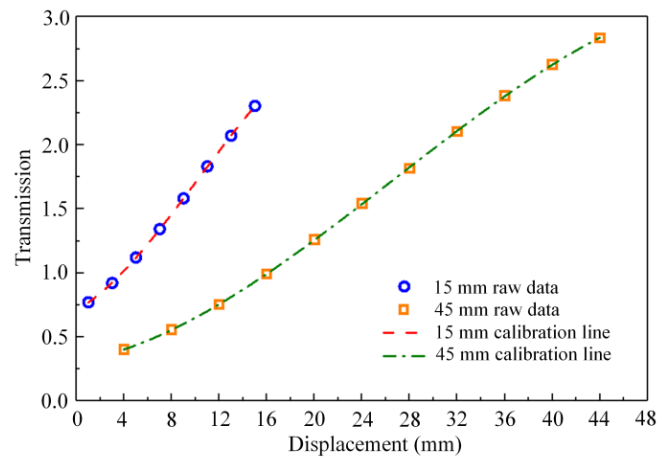

Fig. 7. Transmission Calibration lines with different dynamic range using 200 μs averaged data

TABLE Ⅱ
COEFFICIENT OF CALIBRATION EQUATION

| Coefficient | 15 mm calibration equation | 45mm calibration equation |
| --- | --- | --- |
| $a$ | 0.05291 | 0.01594 |
| $b$ | 0.00690 | 0.00211 |
| $c$ | $-2.3166 \times 10^{-4}$ | $-2.6388 \times 10^{-5}$ |
| $d$ | 0.70997 | 0.30690 |

Once the mapping relation between the displacement and transmission is established, real-time, ultrafast and continuous displacement measurement can be performed.

The final displacement measurement results with different detection ranges are shown in Fig. 8. At each displacement, 100 times measurements are made with 10 μs averaging. The detection speed of 0.1 MHz is realized. Over 15 mm detection range, standard deviation of 7.64 μm and overall mean error of 19.10 μm are obtained. Over 45 mm detection range, standard deviation of 37.73 μm and overall mean error of 36.63 μm are obtained. The trade-off between detection sensitivity and dynamic range occurs as expected.

In the experiment, we notice that, although all-fiber structure of the system is used, it is sensitive to environment disturbance. Vibrations caused by walking or speaking, airflow from air conditioning, temperature fluctuations caused by instruments nearby, and human thermal radiation are possible to change the displacement under test. For long-time operation, periodic calibration is needed.

TABLE III
COMPARISON OF OUR PREVIOUS AND THIS WORK

| Parameter | Previous work ref [14] | This work with filter 1 | This work with filter 2 |
|---|---|---|---|
| Dynamic range | 8.17 mm | 15 mm | 45 mm |
| Detection speed | 48.6 MHz | 0.1 MHz | 0.1 MHz |
| Standard deviation | 334 nm | 7.64 μm | 37.73 μm |
| Mean error | 85 nm | 19.10 μm | 36.63 μm |
| OE Bandwidth | 7 GHz [a] | 20 GHz [b] | 20 GHz [b] |
| Digitization Bandwidth | 7 GHz | 0.35 GHz | 0.35 GHz |
| Sampling rate | 20 GS/s | 1.25 GS/s | 1.25 GS/s |
| Signal processing | Post-processing | Continuous | Continuous |

[a] The OE bandwidth is the lower value of the PD and the Oscilloscope;
[b] The OE bandwidth is the lower value of the PD1 and the intensity modulator.

Oscilloscope is used in the experiment for convenience, and the data are processed by a laptop computer. In practice, an acquisition card with lower sampling rate can be used and data size can be reduced to less than 1 GB per second. So that it is possible to realize real-time data processing by fast FPGA (Field-Programmable Gate Array) card. In table III, our previous work is compared with this work, the dynamic range is increased whereas standard deviation and mean error are increased. In this work, the bandwidth requirement for digitization is reduced from 20 GHz to 0.35 GHz. Although the measured target is static in the demonstration, the proposed method can be used for fast target measurement due to femtosecond detection regime.

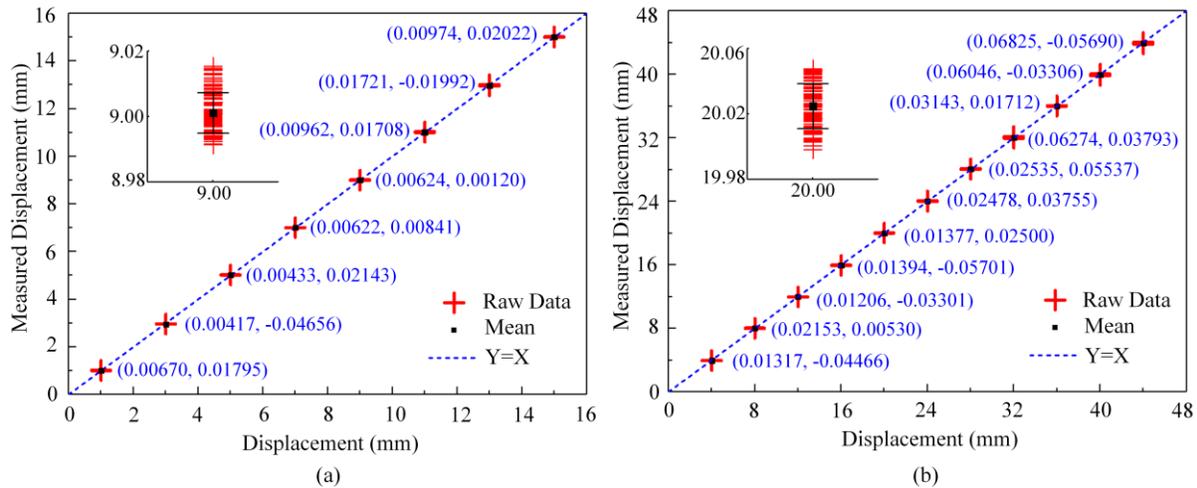

Fig. 8. (a). 15 mm range 10 μs averaged results (b). 45 mm range 10 μs averaged results. There are 100 times measurements at each displacement. Data are shown in format of (standard deviation, mean error)

## IV. CONCLUSION

A femtosecond lidar with tunable dynamic range based on dispersive Fourier transformation and microwave photonic signal processing is experimentally demonstrated. Real-time displacement-frequency-transmission mapping are built. The data size can be tremendously reduced to achieve real-time data storage and processing. According to different applications requirements, the detection sensitivity is tunable by changing the bandwidth of programmable filter. A higher sensitivity can be achieved with a sacrificed dynamic range. As the coherent detection is adopted in this work, it can approach quantum limit sensitivity if the noise of PD1 is dominated by the shot-noise of the heterodyne photocurrent. For long range detection in the future work, a large-area telescope and high-power laser should be used to guarantee the signal to noise ratio.